# Hyper-velocity impact test and simulation of a double-wall shield concept for the Wide Field Monitor aboard LOFT

E. Perinati[a,*], M. Rott[b], A. Santangelo[a], S. Suchy[a], C. Tenzer[a], E. Del Monte[c], J.-W. den Herder[d], S. Diebold[a], M. Feroci[c], A. Rachevski[e], A. Vacchi[e], G. Zampa[e] and N. Zampa[e]

[a]IAAT - Institut für Astronomie und Astrophysik, Universität Tübingen, 72076 Tübingen, Germany
[b]LRT - Lehrstuhl für Raumfahrttechnik, Technische Universität München, 85748 Garching, Germany
[c]INAF - Istituto di Astrofisica e Planetologia Spaziali, 00133 Rome, Italy
[d]SRON- Netherlands Institute for Space Research, 3584 Utrecht, Netherlands
[e]INFN - Istituto Nazionale di Fisica Nucleare/Sez. Trieste, 34127 Trieste, Italy

**ABSTRACT**

The space mission *LOFT* (*Large Observatory For X-ray Timing*) was selected in 2011 by ESA as one of the candidates for the M3 launch opportunity. *LOFT* is equipped with two instruments, the Large Area Detector (LAD) and the Wide Field Monitor (WFM), based on Silicon Drift Detectors (SDDs). In orbit, they would be exposed to hyper-velocity impacts by environmental dust particles, which might alter the surface properties of the SDDs. In order to assess the risk posed by these events, we performed simulations in ESABASE2 and laboratory tests. Tests on SDD prototypes aimed at verifying to what extent the structural damages produced by impacts affect the SDD functionality have been performed at the Van de Graaff dust accelerator at the Max Planck Institute for Nuclear Physics (MPIK) in Heidelberg. For the WFM, where we expect a rate of risky impacts notably higher than for the LAD, we designed, simulated and successfully tested at the plasma accelerator at the Technical University in Munich (TUM) a double-wall shielding configuration based on thin foils of Kapton and Polypropylene. In this paper we summarize all the assessment, focussing on the experimental test campaign at TUM.

**Keywords:** meteoroids, debris, ESABASE2, dust accelerators, Whipple shield, SDD, *LOFT*

## 1 INTRODUCTION

The space environment contains dust, i.e. meteoroids and orbital debris. Meteoroids are natural interplanetary particles with mostly cometary and asteroidal origin, while fragments originating from manmade objects placed in orbit are referred to as debris. Meteoroids and debris, whose typical size ranges from less than 1 μm to above 1 cm, travel at hyper-velocities. Therefore, impacts by spatial dust represent a potential threat for space-borne instruments, due to the structural damages they can produce, which in turn may imply degradation of the performance or even, in the worst case, the failure of the instruments. As an example, both MOS and pn cameras on-board *XMM-Newton* suffered a number of impacts, which caused significant damages [1] and even the failure and loss of CCD#6 (on 9 March 2005) [2] and CCD#3 (on 11 December 2012), both belonging to MOS1; on 27 May 2005 an impact caused a severe damage to the MOS CCD of the X-RayTelescope (XRT) on-board *Swif*t, which resulted in the inability to operate in all modes [2]. In these cases, it is believed that hyper-velocity dust particles entered the Field-Of-View (FOV) of the telescope being focussed by the telescope or being fragmented after impacting against the mirror shells, generating a spray of secondary fragments, some of which were scattered down to the focal plane. In non-focused instruments the resulting damage might be even more serious, as particles can more easily reach the detectors without any loss of kinetic energy along the way. The configuration of the LAD [3] and WFM [4] instruments aboard the space mission *LOFT* [5], which are based on collimated or coded aperture SDDs where electrodes implanted at the surface are used to achieve a very fast drift and collection of the charge produced by photons, suggests that hyper-velocity impacts are an important hazard factor. The assessment of the impact risk is crucial to estimate the probability of damages, quantify the degradation of the instrumental performance, and possibly identify and implement suitable shielding solutions. In this paper we briefly present *LOFT* (Section 2) and the dust environment in its orbit (Section 3), from which we derive an estimate of the number of impacts expected over the mission duration. In section 4 we explain that the risk of performance degradation would be considerable for the WFM and illustrate the risk mitigation strategy: we designed, simulated and tested a possible double-wall shielding configuration that represents a good compromise between reducing the impact risk and worsening the quantum efficiency of the instrument at the lowest energies. Finally, in Section 5 we discuss the case of the LAD and argument that for this instrument the risk of performance degradation is low.

_________________________________________________

*Email: Emanuele.Perinati@uni-tuebingen.de;  Tel.: +49 7071 29 73457

## 2. THE *LOFT* SPACE MISSION

*LOFT* [6] is a medium-class space project proposed as part of the ESA Cosmic Vision program. In 2011 it was selected by ESA as one of the candidates for the M3 launch in 2022-2024, with a nominal duration of 3 (+2) years. It would be placed in a low-Earth near-equatorial orbit (inclination < 5°) at ~600 km altitude. The scientific goal of *LOFT* is the investigation of the strong-field gravity and the equation of state of the ultradense matter, through the observation at high count-rate of compact objects in the 2 to 30 keV energy band with an unprecedented effective area (~10 $m^2$ @8 keV). The adopted innovative technology relies on the usage of SDD tiles [7] derived from those developed by INFN for the ALICE experiment at LHC/CERN. Each tile is a 450 μm thick fully depleted Silicon wafer with a quite large size (~10x7 cm) and capable of achieving high temporal resolution (~7 μsec) combined with good spectral resolution (<260 eV FWHM). A system of cathodes implanted at the surface, on both sides, provides a potential varying from the center of the detector towards the two opposite edges, allowing for a drift of the charges generated by the absorbed photons towards a set of collecting anodes (Figure 1).

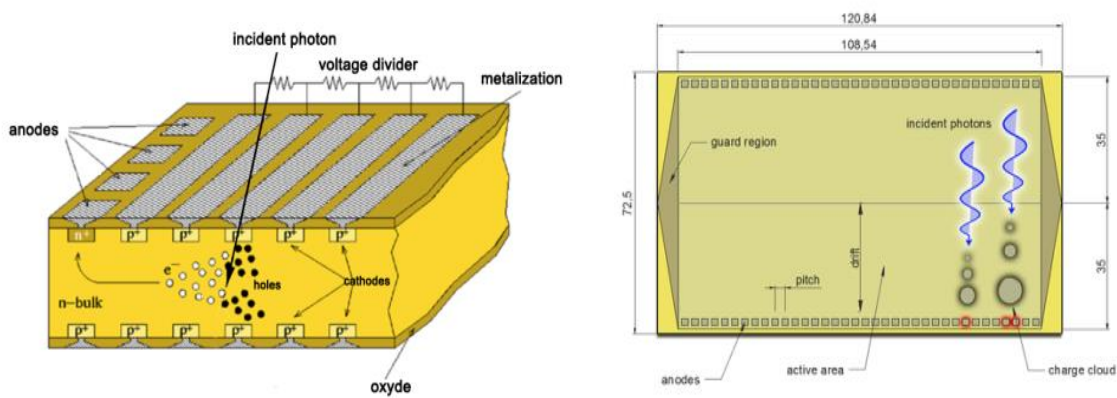

Figure 1. Schematic of the SDD used for LOFT (credit: ISDC). The reported dimensions are typical values of the prototypes produced up to now and do not represent necessarily the final ones.

There are two instruments onboard *LOFT* (Figure 2): the Large Area Detector (LAD) [3] and the Wide Field Monitor (WFM) [4]. In the design proposed by the consortium, the LAD features a large geometric area of ~15 $m^2$ achieved by assembling 2016 SDD tiles into 126 individual modules (4x4 tiles each) divided on 6 panels (3x7 modules each). Other configurations based on a minor number of panels have been investigated as well. It is a collimated experiment and the

6 mm thick micro-pored glass collimator on top of the SDD assembly has a fov of ~$10^{-4}\pi$. The WFM consists of 10 cameras divided in 5 pairs, each camera is a coded-mask experiment with 4 SDD tiles in the detection plane. The WFM has an entrance area of ~0.7 m$^2$ and a large FOV of ~ 5.5 sr (at zero-response), useful to enhance the observation of variable sources.

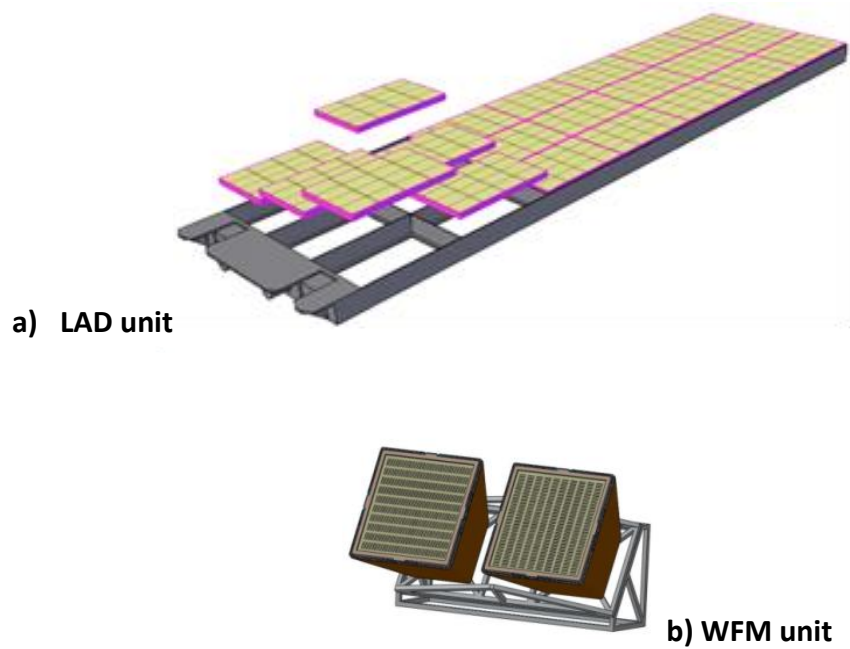

**a) LAD unit**

**b) WFM unit**

Figure 2. a) LAD unit:1 panel = 21 modules = 16x21 SDDs; b) WFM unit: 2 cameras = 4x2 SDDs (credit: ISDC)

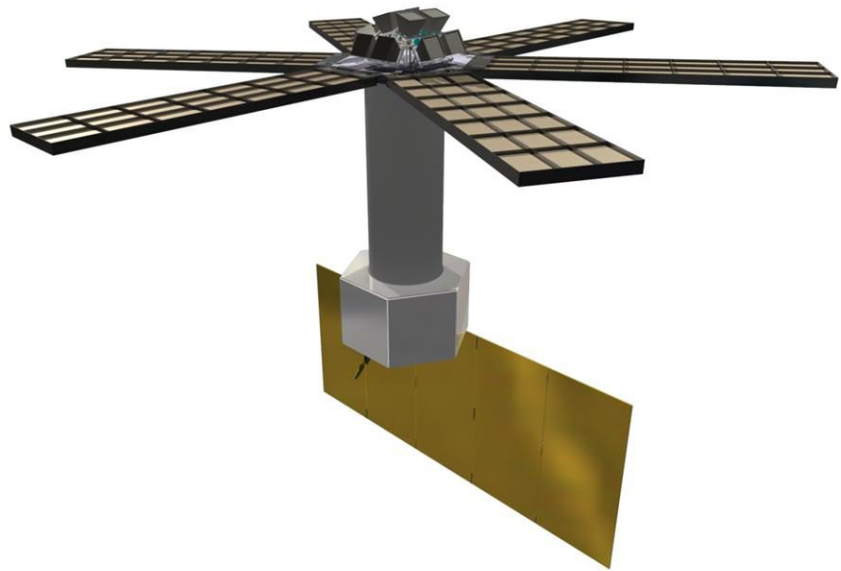

Figure 3. The *LOFT* geometry with the 6 LAD panels and the 10 WFM cameras on top.

## 3. DUST ENVIRONMENT

### 3.1 Meteoroids and orbital debris populations in the *LOFT* orbit

Meteoroids are particles with natural origin that are present in the interplanetary space, originating generally from comets or asteroids. Meteoroids smaller than 1 mm are usually referred to as micrometeoroids. The precession of a satellite orbit and the tilt of the Earth´s equatorial plane with respect to the ecliptic plane implies that the flux of meteoroids can be assumed isotropic relative to Earth. According to [8], the smallest micrometeoroids (<10 μm) include Fe particles (~46%), silicate particles (~17%) and particles made of a lower density material (~37%). At larger size (>10 μm) most meteoroids are expected to be silicate particles, with an average density of ~2.5 g/cm$^3$. On the other hand, orbital debris consists of fragments of various size left in the space environment from manmade objects. Unlike meteoroids, debris is not isotropically distributed and its flux depends on the altitude and inclination of the orbit, with highly inclined orbits usually presenting higher concentration of debris. Furthermore, along an orbit the concentration of debris can be strongly dependent on the direction of observation. The smallest debris (<20 μm) include mostly alumina particles generated when solid rocket motors burn out, while at larger size silicate particles are the most abundant ones [9]. Meteoroids and orbital debris are gravitationally accelerated to hyper-velocities and therefore represent a big hazard for instruments placed in orbit. The average velocity is ~20 km/sec for meteoroids and ~10 km/sec for orbital debris. Figure 4 shows the expected fluence of micrometeoroids and debris in the *LOFT* orbit, according to the cumulative populations reported in [10].

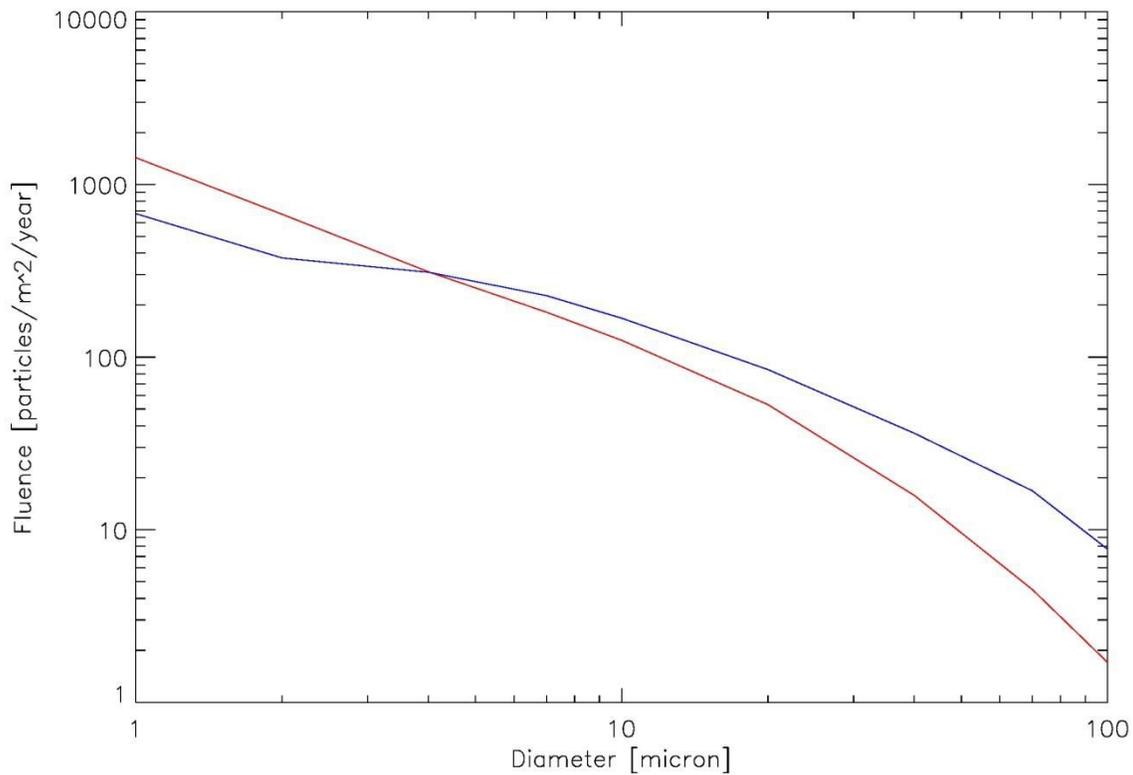

Figure 4. Average cumulative fluence of micrometeoroids (*red line*) and debris (*blue line*) expected in the *LOFT* orbit for a randomly oriented plate in the *LOFT* orbit [10].

## 3.2 Impact effects

When a hyper-velocity impactor hits a target, the resulting structural damage is the formation of a crater on its surface (or of a clear hole in case the impactor has enough energy to fully penetrate through the target). The impact produces a partial vaporization of the impactor with associated ionization and the formation of a plasma cloud, that rapidly expands from the interaction point. This plasma cloud is fundamental in hyper-velocity experiments on ground, in fact placing an electric field around a target and monitoring the variations induced by the released charge provides a trigger signal for the impacts, as we will describe in Section 4.

The morphology of the crater is somewhat different for ductile and brittle materials: in the first case the crater is nearly spherical, while in the other case typically an inner crater with smaller diameter forms within the larger outer crater. The damage equation allows to predict the crater size and depth as a function of a number of parameters. For single-wall targets the general equation for the crater depth (p) is [11]:

$$p = c \cdot d_p^{\alpha} \cdot v_p^{\beta} \cdot \rho_p^{\gamma} \cdot \rho_t^{\delta} \cdot \cos^{\varepsilon} \Theta \qquad (1)$$

where $d_p$, $v_p$ and $\rho_p$ are diameter, velocity and density of the impacting particle, $\Theta$ is the angle of incidence, $\rho_t$ is the density of the target, and c is a characteristic constant. This empirical formula has a general validity, however the values of the parameters $\alpha, \beta, \gamma, \delta$ and $\varepsilon$ are somewhat different for ductile and brittle materials and also vary depending on the range of the variables. Equation (1) and commonly used values of the parameters are reported in ESABASE2. The expected effect of hyper-velocity impacts on the SDD tiles on-board *LOFT* is a permanent increase of the anode leakage current level, and a possible alteration of the electric field at the surface caused by craters. The SDD surface has a $SiO_2$ passivation layer with a thickness of ~1 µm on top, that works as a protection for the underlying active region, only hyper-velocity particles with enough energy to penetrate through the $SiO_2$ reaching the depleted bulk are expected to affect the functionality. By adopting the commonly used Cour-Palais set of parameters for brittle materials (c=1.06, $\alpha$=1.06, $\beta$=0.67, $\gamma$=0.50, $\delta$=0.00, $\varepsilon$=0.67), equation (1) predicts, for example, that even a silicate particle as small as 1 µm impacting at 10 km/sec may be dangerous for the SDD, producing a crater ~2.2 µm deep on its surface for an impact at vertical incidence. In the following Sections 4 and 5 we present the impact risk assessment for the *LOFT* instruments, starting from the case of the WFM, where, due to the much larger FOV, the impact probability results notably higher than for the LAD.

## 4. IMPACT RISK ASSESSMENT FOR THE WIDE FIELD MONITOR

Each WFM camera has a Tungsten-based coded-mask with geometric area A~0.07 m$^2$ and a zero-response FOV~0.8·$\pi$. The open fraction is OF=0.25. The number of impacts at the end of the mission is calculated as:

$$n_{imp} = F \cdot A \cdot OF \cdot FOV \cdot T / 2\pi \qquad (2)$$

where F is the flux of dust particles and T is the duration of the mission. We do not take into account possible stray-dust that might reach the SDDs from directions off-FOV, after hitting the coded-mask walls. We report in Table 1 the average number of impacts by meteoroids and debris in 5 years resulting from (2).

Table1. Average number of meteoroids and orbital debris of a given size expected for each single WFM camera in 5 years. We assumed a geometric area A=0.07 m² (with an open fraction OF=0.25) and a FOV=0.8·π.

| Particle size | Meteoroids in 5 years | Debris in 5 years |
|---|---|---|
| ≥1,≤ 10 µm | ~46 | ~18 |
| ≥10,≤50 µm | ~4 | ~6 |
| ≥50,≤ 100 µm | ~0.3 | ~0.7 |
| >100 µm | ~0.06 | ~0.3 |

For each camera, in 5 years we predict a number of impacts by particles (micrometeoroids and debris) smaller than 10 µm between ~51 and ~79, and a number of impacts by particles in the range 10 to 100 µm between ~5.4 and ~16.9, to a 0.9 CL. While impacts by the smaller particles are expected, at worst, to degrade the performance of the SDDs, as we briefly discuss in Section 5, impacts by particles larger than 10 µm may determine their complete failure. Therefore, assuming a Poisson distribution of the events, and considering that simulations show that particles up to ~10 µm would likely be stopped by the combined effect of the ~8 µm thick Kapton optical/thermal filter, mounted on the coded-mask, and the $SiO_2$ passivation layer on top of the SDD depleted volume, an average value of ~10 impacts corresponds to a probability of ~99 % that all four SDDs are lost. Clearly this probability is unacceptable and additional shielding is required to mitigate the risk.

### 4.1 Shield design

The shield for the WFM should be able to stop hyper-velocity particles up to a size of ~100 µm. This may be achieved with a Whipple configuration (Figure 6), using an outer layer (acting as a *bumper*) plus an inner layer (acting as a *rear-wall*) made of a material capable to withstand hyper-velocity impacts and placed at a certain distance behind the first one. The bumper is expected to shock the primary impacting particle, converting part of the kinetic energy into heat, which results in an explosive fragmentation producing a spray of much smaller grains. The cloud of secondary, less energetic, fragments emerging from the bumper has to be fully absorbed in the rear-wall layer. For the WFM, the challenge is to obtain the desired performance with layers as thin as possible, in order to minimize the reduction of the quantum efficiency of the instrument. The distance between the two layers is approximately 20 cm, corresponding in the current baseline geometry to the distance between the coded-mask and the detector plane.

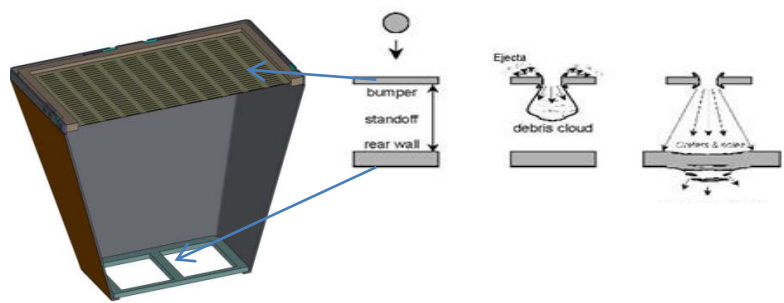

Figure 6. A Whipple shield is composed by an outer bumper layer placed at a certain distance in front of a rear-wall. The bumper produces fragmentation of the primary impacting particle, the spray of less energetic secondary fragments is then stopped by the rear-wall layer. Arrows indicate where the bumper and rear-wall layers would be placed in each WFM camera (half camera is shown here).

The bumper for the WFM consists of a layer of ~8 µm thick Kapton, which is already envisaged in the WFM configuration as an optical/thermal filter. Beryllium is the baseline material for the rear-wall, thanks to its high transmissivity in the soft X-ray band: a 25 µm thick Beryllium plate has been proposed to preserve as much as possible the quantum efficiency of the instrument at the lowest energies. Even though there are a number of models to predict the performance of a Whipple shield, each can be assumed valid only in the usually very restricted set of configuration parameters in which was derived and validated. As there is a lack of experimental data about the response of thin layers to hyper-velocity impacts, we considered it necessary to perform a laboratory test on the shielding concept described in the previous Section. However, due to the well known toxicity of Beryllium, a Beryllium plate is not suited for laboratory tests that generates dust fragments. We searched for an alternative usable in laboratory, and found that Polypropylene $(C_3H_6)_n$, provides a high mechanical resistance as well as a good X-ray trasmissivity: 15 μm Polypropylene have the same X-ray trasmissivity as 25 μm Beryllium (Figure 7).

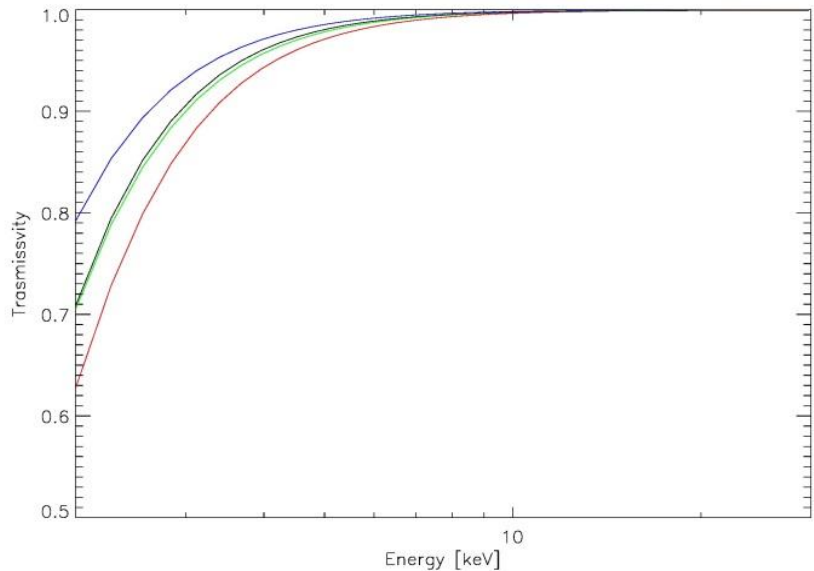

Figure 7. Trasmissivity of a 25 μm thick Beryllium foil (*black line*), 20 μm thick Polypropylene foil (*red line*), 15 μm thick Polypropylene foil (*green line*) and 10 μm thick Polypropylene foil (*blue line*) in the WFM band (from CXRO). The trasmissivity curves of a 25 μm thick Beryllium foil and of a 15 μm thick Polypropylene foil are perfectly overlapped.

### 4.2 Hyper-velocity test campaign at TUM

#### 4.2.1 Plasma accelerator

A hyper-velocity impact test campaign was conducted at the plasma accelerator at TUM [12]. Figure 8 shows a schematic description of the accelerator. The discharge of a 16 kV capacitor bank through ignitron switches into a plasma gun filled with a few cm$^3$ of Helium gas generates a plasma, which is accelerated along the coaxial electrodes by electrical currents and magnetic fields. The plasma reaches the compressor coil at about 80 km/sec, where the flow is compressed and the particles exposed to the flow pressure are drag accelerated. This machine is unique in the generation of hyper-velocity particles of intermediate size, as Van de Graaff accelerators can accelerate only smaller particles of a few micron size and light gas guns usually work with particles larger than 100 μm. The typical particle distribution achievable at TUM is in a range from 10 μm to approximately 100 μm size. The chemical species used in the accelerator is silicate with a density of 2.5 g/cm$^3$, which is expected to simulate well most of the micrometeoroids and orbital debris in this size range. The accelerated particles hit the target at nearly vertical incidence. This represents a conservative simulation of the orbital case, as in orbit oblique impacts, which are expected to produce less damage than vertical ones, are more probable due to the large FOV of the WFM cameras.

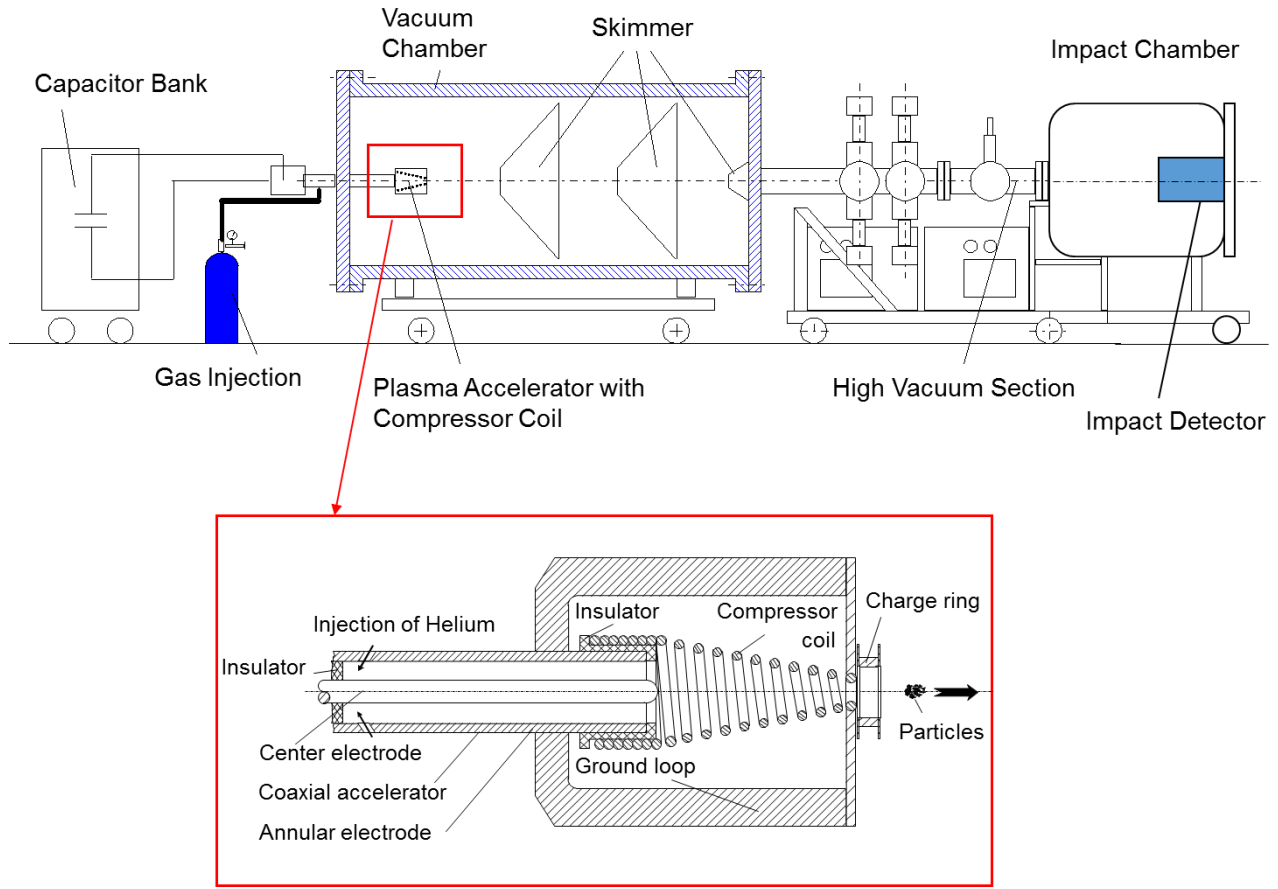

Figure 8.   Schematic of the plasma accelerator at TUM

**4.2.2   Experimental setup**

We built several individual targets with the same specifications in order to shoot only few particles against each, for an easier inspection of the layers in the optical microscope after the bombardment. Each target is a shield sample made of a layer of 8 μm thick Kapton coupled to a layer of 15 μm thick Polypropylene. To exploit the mechanical mounting that already existed at the accelerator we placed the Kapton bumper and the Polypropylene rear-wall at 4.5 cm distance from each other, the largest possible value allowed in that configuration (Figure 9). This spacing is shorter than the actual distance exploitable on the WFM, i. e. our experimental configuration is less effective and the results of the test can be assumed as conservative (we develop this argument in the caption of Figure 15).  A possible issue was how to get a clear understanding of the results of the test, as it would have been difficult to look in the microscope for the very small craters or holes produced by secondary fragments and distinguish them from the defects and impurities on the unpolished Polypropylene layers. To make the inspection easier and faster, we plated the Polypropylene layers with 150 nm Aluminium. This allows to easily locate and recognize the pattern of tiny imprints created by the spray of secondary impacting fragments, as the Aluminium coating makes the layer opaque to the visible light, but in the collision points the impacting fragments remove the metal and the layer becomes transparent. By illuminating it from the uncoated back side with a laser beam or a lamp, it is possible to see clearly the pattern produced by an impact (Figures 15, 16 and 17).

Eventually, a polished steel plate is placed behind the rear-wall to detect possible imprints left by ejecta emerging from Polypropylene as a proof of rear-wall perforation. We applied the same Aluminium coating to the Kapton layers, where the presence of the metal is useful to have enough charge generation during the bombardment for triggering the detection of the impacts. As the applied coating is much thinner than the layers and the particle size, we assume that it did not have any influence on the dynamics of the impacts.

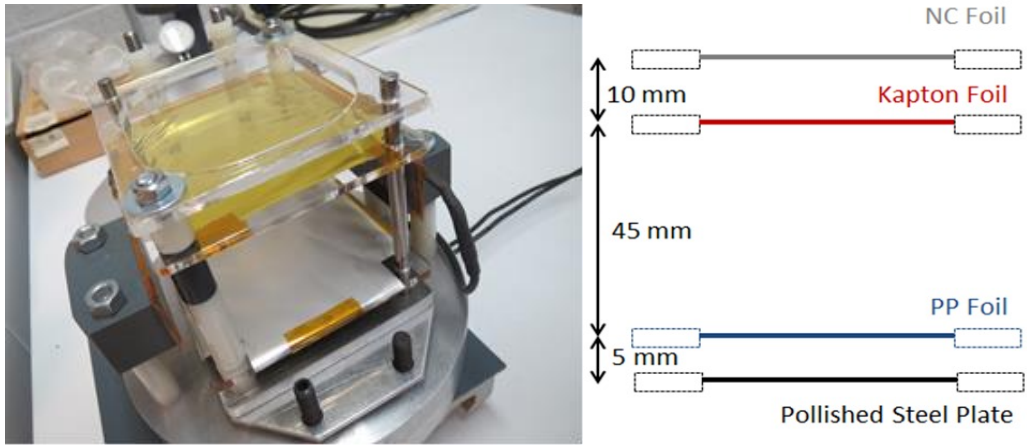

Figure 9. The shield target: all layers are fixed to plastic supporting frames 9x8 cm. At the two sides of the target the two charge collecting plates are visible.

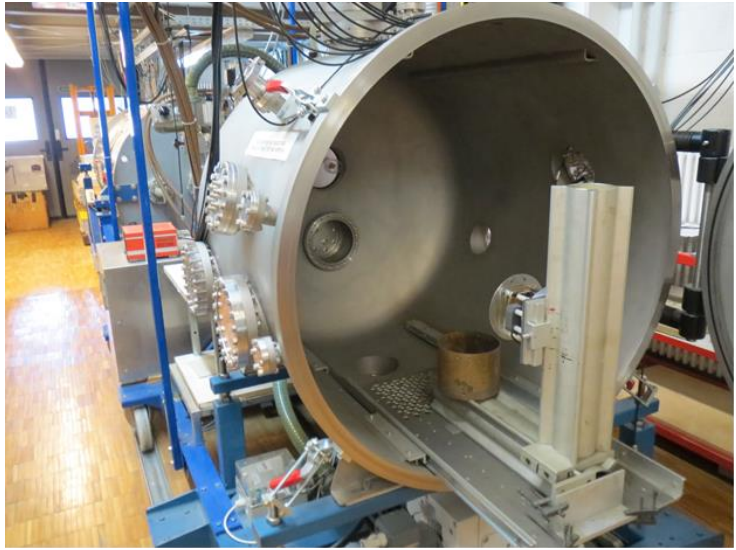

Figure 10. The target mounted in the vacuum chamber. A metal cylindric case (here visible on the bottom of the chamber) is placed around the target and the collecting plates to get rid of the electromagnetic noise generated by the accelerator.

Figure 10 shows the target mounted in the vacuum chamber of the accelerator. Ultrafast drag acceleration leads to ablation and deflection of the particles, therefore it is not possible to select size, velocity and impact point prior to a shot, instead the impact parameters are measured after each shot. Figure 14 explains the details of the procedure: two charge collector plates are mounted at the side of the target, the detection of a charge signal by these plates gives information that an impact occurred. Relating the time of this trigger signal with the time of the capacitor discharge makes it possible to estimate the velocity of the particle, as the distance between the compression coil and the target is known (4.5 m).

There is a short delay between the capacitor discharging time and the actual time the particle leaves the coil, which from previous tests was estimated ~20 μsec, on average. As all particles produced by the accelerator need at least ~450 μsec to cover the distance between the coil and the target, an error of ~20 μsec corresponds to an accuracy in the estimate of their velocity better than 10%. A special nitrocellulose (NC) thin film (100 nm thick) is mounted in front of the target, the impacting particles pass through it without any fragmentation and producing clear holes. Measuring the size of the holes in the optical microscope gives precisely the size of the particles impacting onto the target. The precision in determining the diameter of the holes is 1 μm, however in some cases the hole might be not perfectly round, depending on the actual shape of the particle at the impact point, so that in such cases the hole size is estimated by some other criterion based on the actual shape of the hole in the x,y plane. Since the impacting particles leave the largest holes on the Kapton layer, the location of the impacts is determined by inspecting first this layer in the microscope. Then, the position of a hole on the Kapton layer indicates the position on the NC foil where the clear hole is to be expected and restricts as well the area on the Polypropylene for the search for holes or craters left by secondary fragments. Finally, the position of the area affected by the secondary spray on the Polypropylene localizes the area on the polished steel plate for the search for tiny imprints of possible ejecta indicating whether Polypropylene perforation occurred.

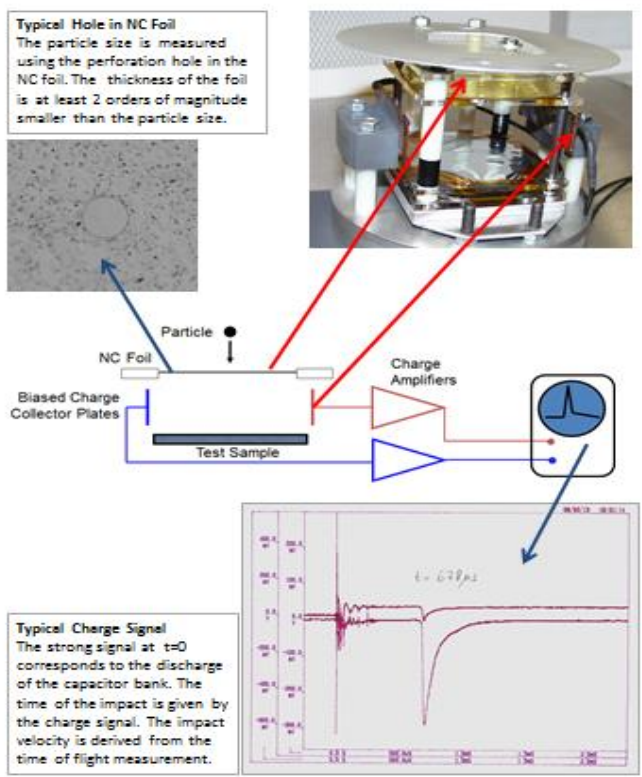

Figure 11. The measurement of the impact parameters

### 4.2.3 Results

During the test campaign we have performed 13 shots for the combination Kapton (bumper) 8 μm + Polypropylene (rear-wall) 15 μm. It is worth stressing that the number of shots is constrained by the long time required for the setting of the accelerator as well as the high costs of running it: on average, a whole day is needed to obtain a single useful shot (a hit on the target), moreover it was necessary to open the chamber after each impact breaking the vacuum in order to remove the target and perform inspection of the bombarded target at the microscope. However, the achieved statistics was enough to understand the physical processes and the response of the shield to hyper-velocity impacts. We observed three regimes of velocity, each associated with a typical behaviour of the particles hitting the Kapton bumper:

a) v>7 km/sec: particles undergo a complete explosive fragmentation or even evaporation (the smallest ones), producing spray of tiny secondary fragments which are then stopped by the Polypropylene rear-wall (Figure 12);

b) 5 <v< 7 km/sec: particles undergo a partial fragmentation, resulting in a number of smaller grains some of which are still big enough to make clear holes in the Polypropylene rear-wall (Figure 13);

c) v < 5 km/sec: particles do not undergo fragmentation and perforate both layers (Figure 14).

Figures 12,13 and 14 have been selected as typical examples that illustrate the scenario described above, the photographs were taken during inspection of the bombarded layers at the microscope.

We compared our results with the predictions of the ESABASE2 International Space Station (ISS) model applied to our configuration and have found that they are in a remarkable good agreement, as shown in Figure 18. We could not validate the model predictions for particles larger than ~70 μm and faster than ~7 km/sec, as that is close to the limit of the accelerator. However, the good agreement in the achievable range of parameters suggests that the model predictions should be expected consistent for different values of the parameters as well. Therefore, for a configuration with 20 cm spacing we conclude that the shield would be able to stop particles up to ~100 μm at 10 km/sec, and particles up to ~60 μm at 20 km/sec. As ~0.2 meteoroids larger than 60 μm and ~0.3 debris larger than 100 μm are expected over 5 years in orbit, we calculate a probability of ~60 % of having no hits in 5 years, assuming a worst case vertical impact occurrence. Considering that within an acceptance angle of $0.8 \cdot \pi$ oblique incidence is more probable than vertical incidence the actual picture is even more favorable, at 30° incidence the probability of having no hits can be expected higher than 85 %.

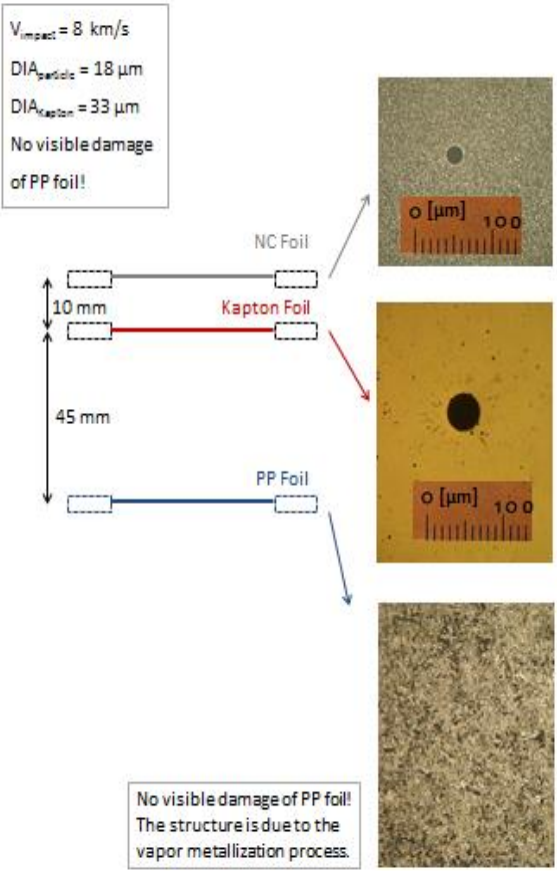 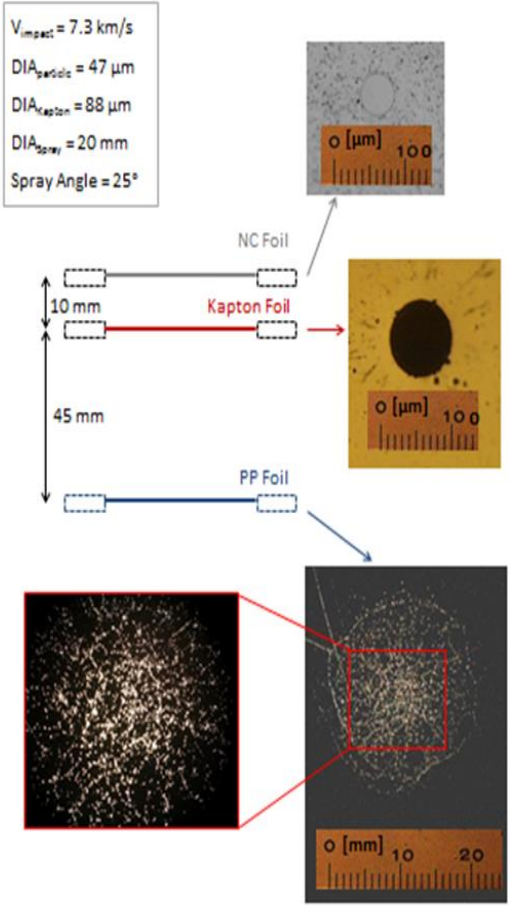

Figure 12. Typical impacts at high velocity: (*left*) the primary particle has a small size and is evaporated by the Kapton bumper, no solid fragments emerge; (*right*) the primary particle has a large size and is completely fragmented by the Kapton bumper producing a spray of secondary tiny (micron and sub-micron sized) fragments fully stopped by the Polypropylene rear-wall.

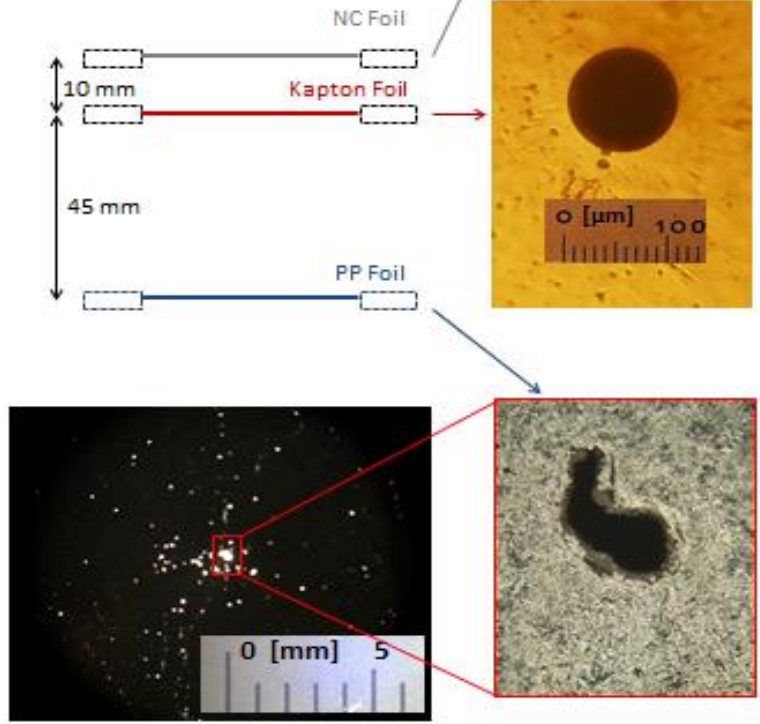

Figure 13. Typical impact at medium velocity: the primary particle is partially fragmented by the Kapton bumper into a certain number of secondary smaller fragments, some of them are still big enough to make holes in the Polypropylene rear-wall (notice that the size of the holes in the photograph appear magnified with respect to the actual one due to the fact that the impacting fragments remove some metallization also around the holes)

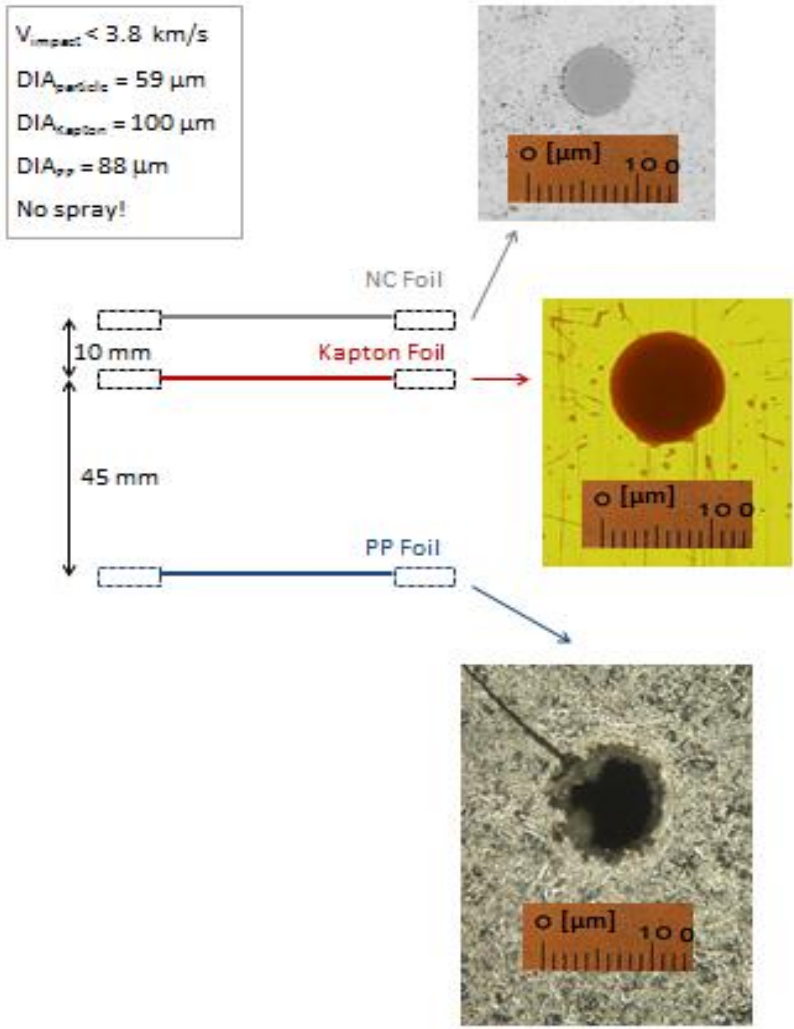

Figure 14. Typical impact at low velocity: the primary particle is not fragmented by the Kapton bumper and it perforates also the Polypropylene rear-wall

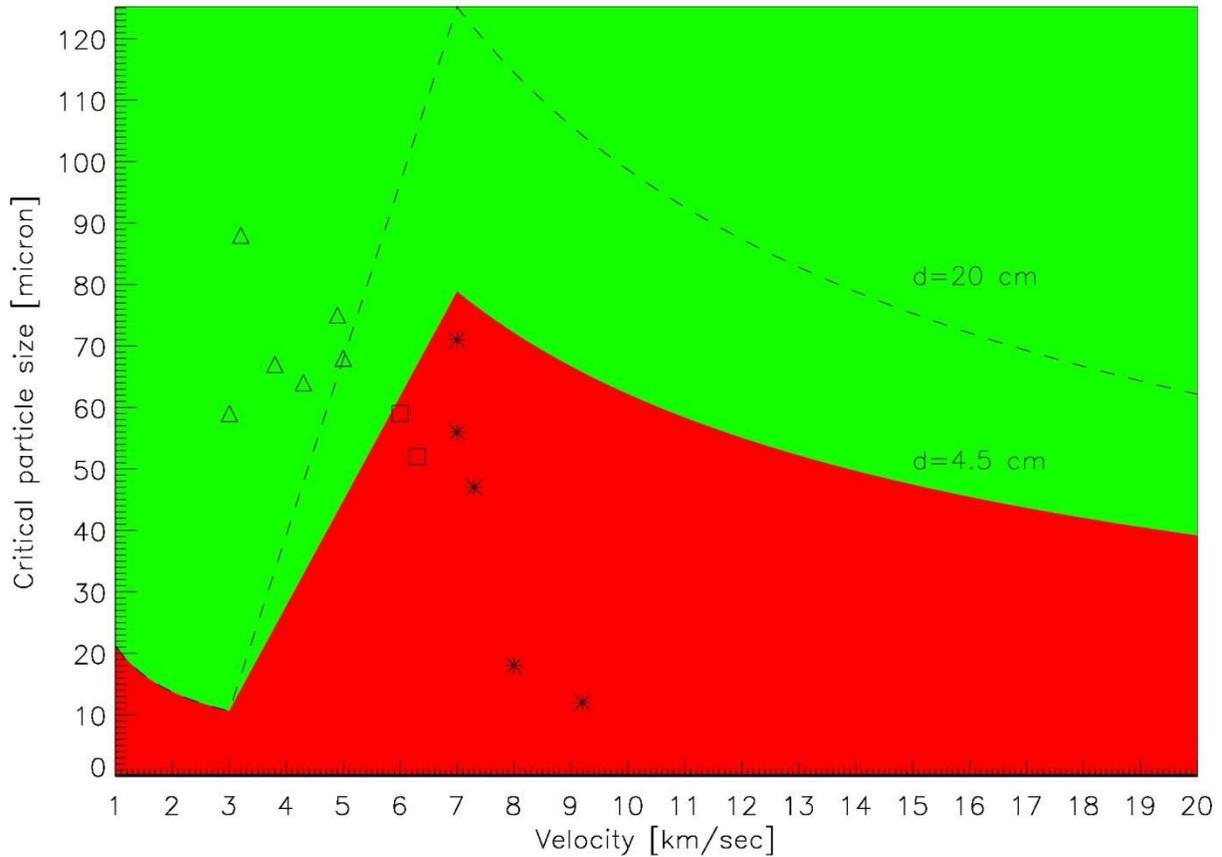

Figure 15. Simulated performance of our Kapton+Polypropylene shield by adopting the ISS model: the red area indicates particle stopped, i.e. the combination of particle parameters for which the shield is effective; the green area indicates particle passing through, i.e. shield failure. The points are experimental data of shots performed in the test campaign at TUM: ∗ = shield effectiveness (Polypropylene was not perfored); □ = unclear case (Polypropylene was perfored but no impact imprints were detected on the polished steel plate below it, possible complete evaporation of the ejecta by Polypropylene); Δ = shield failure (Polypropylene was clearly perfored). In the simulation we assumed 50 MPa as the stress yield of Polypropylene, the point P = (~7 km/sec, ~70 μm) demonstrates that the actual stress yield of our Polypropylene sample might be larger than 50 MPa but should not be much lower. The simulation considers vertical incidence of the particles. The dashed line is a simulation showing where the red area would be extended by increasing the distance between the layers from 4.5 to 20 cm, that corresponds to the actual WFM configuration: a larger distance allows for a larger opening of the spray cone emitted by the Kapton bumper, resulting in the momentum of the secondary fragments impinging onto the Polypropylene rear-wall being distributed over a wider area, then producing a lesser amount of damage.

## 5. IMPACT RISK ASSESSMENT FOR THE LARGE AREA DETECTOR

Despite the huge effective area, the relatively small fov of the LAD restricts the risk of impact only to particles with size less than 10 μm, the most abundant ones in the environment. In Table 2 we report the average rate of particles of a given size expected for the LAD over 5 years in orbit, as it results from (2). Taking into account both micrometeoroids and orbital debris the resulting average rate is ~5 impacts in 5 years.

Table2. Average rate of impacts by micrometeoroids and orbital debris of a given size expected for the whole LAD in 5 years. We assumed a geometric area A = 15 m$^2$ (with an open fraction OF=0.7) and a fov=10$^{-4}\pi$

| Particle size | Meteoroids in 5 years | Debris in 5 years |
|---|---|---|
| $\geq 1, \leq 2$ μm | ~2 | ~0.8 |
| $\geq 2, \leq 3$ μm | ~0.6 | ~0.08 |
| $\geq 3, \leq 4$ μm | ~0.3 | ~0.07 |
| $\geq 4, \leq 5$ μm | ~0.15 | ~0.06 |
| $\geq 5, \leq 6$ μm | ~0.1 | ~0.05 |
| $\geq 6, \leq 7$ μm | ~0.08 | ~0.04 |
| $\geq 7, \leq 8$ μm | ~0.06 | ~0.03 |
| $\geq 8, \leq 9$ μm | ~0.05 | ~0.02 |
| $\geq 9, \leq 10$ μm | ~0.03 | ~0.01 |

However, an optical/thermal filter made of a foil of 1 μm thick Kapton plated with 40 nm Aluminum on both sides is envisaged in the baseline configuration of the LAD. This filter can be mounted above or below the collimator, and the first option would be preferable in terms of protection of the SDDs from dust impacts. In fact, the Kapton+SiO$_2$ system actually works as a double-wall shield with 6 mm spacing. We estimated the shielding effectiveness of this double-wall using the ISS model, assuming 1 μm thickness of SiO$_2$. According to our simulations micrometeoroids smaller than ~3 μm and debris smaller than ~5 μm would not be able to reach the active region below the passivation layer. Therefore, from the data in Table 1 we have that the average number of hazardous impacts over 5 years in orbit reduces to ~1. Assuming a Poisson distribution of the events and considering that the LAD is composed by 2016 indipendent tiles, an average rate of ~1 event corresponds to a probability of ~37 % that the effective area of the LAD is degraded by 0.05 % over 5 years in orbit, ~19 % probability that it is degraded by 0.1 % and virtually no probability that it is degraded by more than 1%. Then, even in the worst case, i.e. assuming that each impact implies a complete failure of the hit tile, this scenario is indeed not concerning. Furthermore, we characterized the response of SDD prototypes to impacts, performing hyper-velocity impact tests at the MPIK dust accelerator in Heidelberg to quantify to what extent the detector functionality might be degraded by the collision with micron-sized dust particles. A detailed discussion of the results of these measurements can be found in [13]. In summary, the laboratory tests evidenced the robustness of the SDD in the range of impact parameters expected for the LAD. Figure 16 is an example of what typically happens in the case of impacts that produce craters deep enough to reach the depleted bulk: the level of leakage current of the SDD undergoes a small step of increase as a consequence of each impacting particle penetrating through the passivation layer. Since impacts produce permanent damages, after a step of increase the leakage current does not recover back to the previous value.

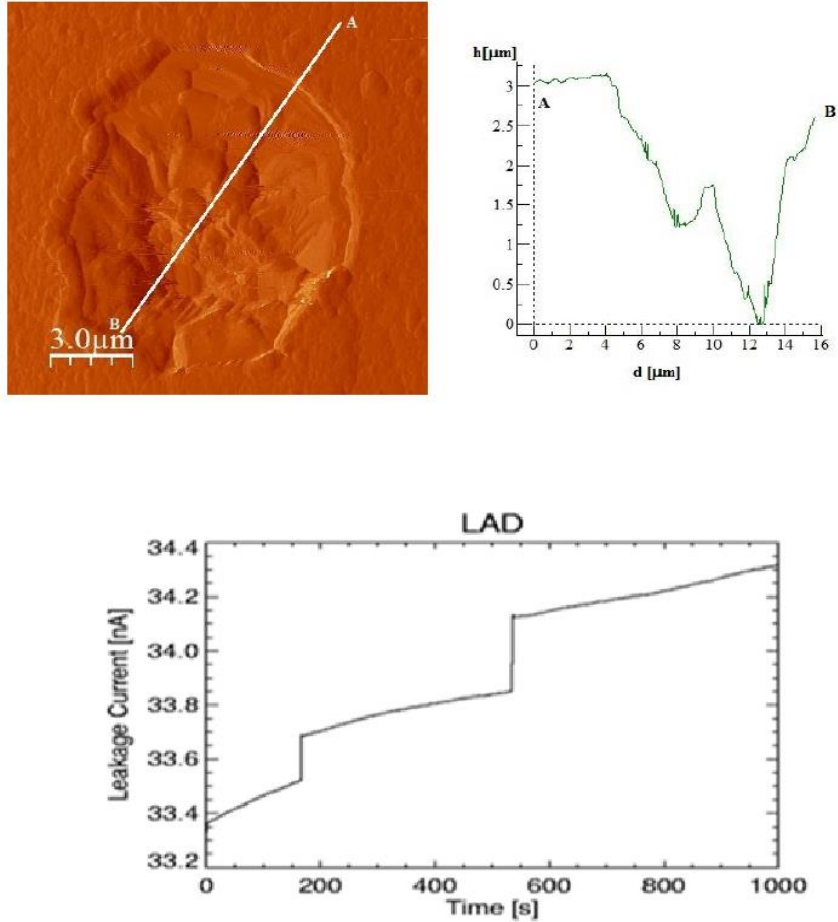

Figure 16. Some results from the impact test campaigns conducted on SDD prototypes at MPIK. *Top:* photograph of a crater produced on the SDD surface by an impact. An atomic force microscope (AFM) was used to measure the crater depth, the vertical profile of the crater along the line from A to B is shown. *Bottom*: transitions of the anode leakage current of the SDD onto higher levels monitored during the bombardment with micron-sized Fe particles. The amplitude of the steps can vary largely even for particles having very similar parameters.

## 6. CONCLUSIONS

We presented an assessment study of the risk posed to the instrumentation aboard LOFT by hyper-velocity impacts. We can summarize our findings as follows:

- for the WFM the risk of damages or failure is considerable, as its large FOV implies that a few relatively big particles may reach the detector plane of the cameras. The optical filter envisaged in the WFM configuration cannot provide enough protection and an additional layer is required to effectively shield the SDDs from these particles and make the impact risk low enough. Placing it just in front of the SDDs at approximately 20 cm distance from the optical filter mounted on the coded-mask realizes a so-called Whipple shield configuration, whose properties have been investigated by simulations and validated by experimental tests conducted at the plasma accelerator at TUM. The result of this study showed that a 15 μm rear-wall made of Polypropylene is capable to effectively stop particles of a few tens micron size, provided that they have a velocity higher than ~5 km/sec enabling a complete disintegration, by means of the Kapton bumper, into a spray of tiny secondary fragments, which then are stopped by the Polypropylene rear-wall. The flux of particles in orbit at velocity lower than 5 km/sec is expected low enough to neglect the risk posed by them. As for the penetration threshold of the shield at high velocity, the maximum momentum we could generate was for particles of ~70 μm size with velocity ~7 km/sec, which is at the limit of the accelerator capabilities. However, in the range of particle parameters obtained from the accelerator our experimental results are in a remarkable agreement with the predictions of the ISS model that we adopted. This fact suggests to rely on such model to estimate the shield performance also at larger sizes and higher velocities, and for different values of the other parameters. The model prediction shows that at ~10 km/sec particles would be stopped up to ~100 μm size, at ~20 km/sec particles up to ~60 μm size. Therefore, according to the expected rate of micrometeoroids larger than ~60 μm and debris larger than ~100 μm reported in Section 3 we estimate a probability of ~60 % (at vertical incidence) or ~85 % (at 30° incidence) that none of the WFM SDDs will be hit over 5 years in orbit. Finally, we stress that the agreement of the ISS model prediction and our experimental data on a sample shield realized with thin plastic foils is remarkable, considering that the model was derived and previously validated in a quite different context (relatively thick layers of metal). This test provides further validation of the model in a range of parameters not well explored by experiments yet;
- for the LAD the risk of a degradation of the effective area due to hyper-velocity impacts is low: only a few particles smaller than 10 μm are expected to hit the LAD SDDs over 5 years in orbit and the optical and thermal filter together with the $SiO_2$ passivation layer on top of the surface can provide some protection. Furthermore, the results of impact tests conducted on SDD prototypes suggest that the detector is quite robust with respect to impacts of micron-sized particles.

## ACKNOWLEDGEMENTS

The authors acknowledge support for the test campaigns at the dust accelerators from the *Bundesministerium für Wirtschaft und Technologie / Deutsches Zentrum für Luft und Raumfahrt (DLR)* under grant FKZ 50 OO 1110. In Italy *LOFT* is funded by ASI under contract ASI/INAF n. I/021/12/010. We thank J. Loebell of the Physikalisches Institut in Tübingen for performing crater inspection on the SDD prototype with their AFM microscope, and Antonio Bonardi for useful discussions and support with the usage of the coating chamber.